\documentstyle[11pt,newpasp,twoside,epsf]{article}
\def\edcomment#1{\iffalse\marginpar{\raggedright\sl#1\/}\else\relax\fi}
\reversemarginpar

\begin{document}
\title{Small-scale structure deduced from X- and $\gamma$-ray timing 
measurements}
 \author{M. Coleman Miller}
\affil{University of Maryland, Department of Astronomy, College Park, MD, USA}

\begin{abstract}
X-ray timing observations of  neutron stars and black holes are among the
few available probes of ultrastrong magnetic fields, strong
gravity, high densities, and the propagation of
thermonuclear burning.  Here we review the evidence for these effects
revealed with data from the Rossi Explorer in the last five years. We
also discuss the exciting prospects for making the first quantitative
tests of strong-gravity general relativistic predictions with
a large-area X-ray timing mission.
\end{abstract}

\section{Introduction}

X-ray timing has historically led to fundamental understanding of many 
phenomena, such as accretion-powered pulsars, low-mass X-ray binaries,
and strongly magnetized neutron stars.  X-ray timing has unique power
because timing signals are often clean diagnostics of
systems, and hence lead directly to physical understanding.
Here we discuss recent results of X-ray timing.  In particular, we
focus on aspects of fundamental physics that can be addressed with timing
observations.  Physics in ultrastrong magnetic fields, high densities,
and strong gravity can all be tested with such observations, which
reveal truly small-scale structure, from the $\sim 10^6$~cm sizes of
neutron stars and black holes to the $\sim 10^{-13}$~cm radii of
nucleons.  In \S~2 we consider strongly
magnetized neutron stars, and the evidence for them apparent in timing
observations.  In \S~3 we discuss thermonuclear propagation in X-ray bursts,
and what it might tell us about such propagation in other astrophysical
settings, such as Type~Ia supernovae.  In \S~4 we examine what has been
learned about strong gravity and dense matter through X-ray timing
observations.  Finally, in \S~5 we consider future prospects, and the
exciting qualitative advances in these fundamental subjects that are
expected from observations with a large-area X-ray timing mission.

\section{Strongly Magnetic Sources}

It has long been recognized (e.g., Erber 1966) that in magnetic fields
$B\ga B_c=m_e^2c^3/\hbar e=4.414\times 10^{13}$~G, 
many otherwise insignificant microphysical processes become
important or even dominant.  Such processes include single photon pair
production, photon splitting, and the existence of a vacuum resonance
frequency.   Terrestrial laboratories are unable to reproduce the
combination of high magnetic field and high photon energy or high
electron Lorentz factor  necessary to test the predictions of quantum
electrodynamics in these extreme conditions.  The study of high-field
neutron stars is therefore important for fundamental physics as well as
for the astrophysical understanding it brings.

Observations of soft gamma-ray repeaters provide the best current
evidence for supercritical magnetic fields $B>B_c$.
The first detection of soft gamma-ray repeaters was made in 1979, with
the detection of a giant burst on 5 March 1979 from the source SGR
0529--66 in the Large Magellanic Cloud (Mazets et al.\ 1979)
It was only much later  that this source and two
similar ones (SGR~1900+14 and SGR~1806--20) were recognized as a class
potentially distinct from classical gamma-ray bursts.  These three
sources, a more recently detected fourth member of the class
(SGR~1627--41; Woods et al.\ 1999), and possibly even a fifth
(SGR~1801--23; Cline et al.\ 2000) have long quiescent periods of
relatively steady  $\sim 10^{34-35}$ erg~s$^{-1}$ X-ray emission
punctuated by episodes of bursting behavior. The bursts are typically
short, from $\la 0.1$~s to $\sim 3$~s, and have peak estimated isotropic
luminosities of $L\sim 10^{39-42}$~erg~s $^{-1}$.  The spectra are
quasithermal, with a peak energy at $\sim 20$~keV.

The first suggestion that soft gamma-ray repeaters might be ``magnetars"
with supercritical magnetic field strengths was made by Thompson
\& Duncan (1995, 1996).  The cleanest evidence in favor of the magnetar
hypothesis has come from timing observations.    Modulation of the
persistent X-ray emission from SGR~1806--20 was detected during the
November 1996 active period (Kouveliotou et al. 1998), at a period of
7.5~s and with a period derivative $\dot{P}=2.6\times
10^{-3}$~s~yr$^{-1}$.  Interpreted using the standard dipole braking
model, these values of $P$ and $\dot{P}$ imply a surface dipole field of
$\approx 8 \times 10^{14}$~G, nearly twenty times the quantum critical 
field.  Similarly, the  ${\dot P}=3.3\times 10^{-3}$~s~yr$^{-1}$
spindown of SGR~1900+14 (Kouveliotou et al.\ 1999) plus its 5.2~s period
also implies a surface dipole field of $\approx 8\times 10^{14}$~G in
this source.

The discovery of the high magnetic field inferred from dipole spindown
models of SGR~1806--20 and SGR~1900+14 is a dramatic confirmation 
of the magnetar model.  Alternative models must of course be considered.
For example, Marsden, Rothschild, \& Ligenfelter (1999) have noted that
the environments of SGRs are unusually dense, and have suggested that 
accretion must therefore be reconsidered as the driving mechanism for SGR 
behavior.  However, the nature of the spindown supports strongly the
interpretation that SGRs have supercritical magnetic fields, and hence
are unique laboratories for the predictions of quantum electrodynamics
in ultrastrong magnetic fields.

\section{X-ray Bursts and Thermonuclear Propagation}

Prior to the launch of the Rossi X-ray Timing Explorer (RXTE) it was expected
theoretically (e.g., Joss 1978; 
Fryxell \& Woosley 1982; Nozakura, Ikeuchi, \& Fujimoto 1984; Bildsten
1995) that ignition would happen at some point in the fuel layer and
then spread, so that burning would not be uniform over the entire star.
These expectations were confirmed by the discovery of brightness
oscillations during thermonuclear X-ray bursts from several neutron
stars in low-mass X-ray  binaries (for reviews see Strohmayer, Zhang, \&
Swank 1997; van der Klis 2000). These brightness
oscillations are highly coherent  ($Q=\nu/$FWHM$\sim 1000$; see, e.g.,
Strohmayer \& Markwardt 1999) and are extremely stable in
their frequency from burst to burst in a given source  (Strohmayer et
al.\ 1998).  Their coherence and stability, plus the recent report of a
burst oscillation in the X-ray millisecond pulsar SAX~J1808-3658 
with a frequency equal to the spin frequency (in 't Zand et al. 2000),
establishes that the oscillations occur at the spin frequency or its
first overtone.  The physical picture is that ignition happens at some
spot on the surface, and the burning then propagates away from this
spot.  At a given moment, there is therefore a hot spot on the surface.
Rotational modulation as seen by a distant observer then produces the
observed oscillations.  The linear rotational velocity, $\sim 0.1c$,
is much greater than the $\sim 10^{6-9}$~cm~s$^{-1}$ velocity expected
for thermonuclear propagation, and hence the rotational modulation
provides a series of snapshots of the burning as it propagates.

The study of thermonuclear propagation in X-ray bursts has broad
implications for other propagation phenomena, such as classical novae
and Type~Ia supernovae.  SNe~Ia are especially important to understand
because of their unique value as probes of the expansion of the universe and
of the possible existence of a cosmological constant.   However, these
sources are difficult to model numerically,  due to the large range of
length scales involved (from the $\sim 1$~cm thickness of flame fronts
to the $\sim 10^8$~cm radius of the pre-SN white dwarf).  Observations
are also challenging, because the propagation of burning happens
primarily deep within the white dwarf, and a given source only has one
episode of such propagation in its lifetime.  This means that, even on
such issues as whether the propagation happens as a deflagration (slower
than the sound speed $c_s$ in the unburned material) or a detonation
(faster than $c_s$), and the expected transition between those
propagation regimes, there is considerable uncertainty. Observation of
thermonuclear propagation during X-ray bursts offers a different path by
which these processes can be understood.  The burning happens close to
the surface ($\sim$meters), where it is relatively easily observed, and
happens frequently for a given source, so many observations can be made.
Careful measurement of burst brightness oscillations, especially with a
future high-area timing instrument, therefore holds promise for the
elucidation of thermonuclear propagation in other circumstances such as
Type~Ia supernovae.

One recent result about propagation emerged from study of the burst
source 4U~1636--536, in which the dominant 580~Hz
brightness oscillation is at twice
the spin frequency (Miller 1999). This is interpreted as the existence
of two nearly antipodal hot spots on the surface, perhaps the result of
two pools of fuel (e.g., at the magnetic poles).  In general, ignition
is expected to happen at one pole first, then propagate to the other,
so the rapidity with which the 580~Hz oscillation appears indicates the
speed of propagation.  In the case of 4U~1636--536, the 580~Hz
oscillation appears within 0.03~s of the burst onset.  Combined with
the  $3\times 10^6$~cm between antipodes, this implies a
propagation velocity $v>10^8$~cm~s$^{-1}$, much larger than the 
$\sim 10^6$~cm~s$^{-1}$ expected in some models of turbulent deflagration
waves (see, e.g., Fryxell \& Woosley 1982; Nozakura et al.\ 1984).

These numbers point to one of many qualitative advances that would be
possible with an X-ray timing mission with 10x the collecting area of
RXTE.  The speed of sound in the unburned medium is expected to be
$2-3\times 10^8$~cm~s$^{-1}$.  With the observed signal strength, an
instrument with ten times RXTE's area would be able to resolve propagation
velocities up to $10^9$~cm~s$^{-1}$, and would therefore be able to
distinguish between deflagration and a detonation and possibly see a 
transition between the two burning regimes.  This would
be a major advance in the understanding of thermonuclear propagation in
general.

\section{Strong Gravity and Dense Matter}

Many of the most dramatic advances afforded by X-ray timing have
occurred in the last few years, in the subjects of strong gravity
and dense matter.  This is a direct result of the high area, fast timing,
and high telemetry rate of the Rossi X-ray Timing Explorer.  In this
section we summarize what has been learned from high-frequency
brightness oscillations from neutron stars and black holes, and
in the next section we discuss the dramatic qualitative advances
in our understanding that would be made possible by a future timing
mission with ten times the area of RXTE.

\subsection{Neutron stars}

In addition to brightness oscillations during X-ray bursts (discussed
in the previous section), the major discovery with RXTE was the
existence of kHz quasiperiodic brightness oscillations (QPOs)
in neutron-star low-mass
X-ray binary systems and the existence of a number of QPOs
in accreting black holes.  The QPOs in
neutron stars are seen from $\sim$20 systems, and a number of trends
have been established (see van der Klis 2000 for a review).  
These QPOs (1)~often appear as two simultaneous
peaks in a power density spectrum, (2)~are relatively sharp and strong,
with $Q\equiv \nu/$FWHM up to $\sim 100-200$ in some cases and fractional rms
amplitudes up to 15\% in the 2-60~keV range of the Proportional Counter
Array on RXTE, (3)~tend to increase in frequency as the inferred mass
accretion rate increases, (4)~have a separation frequency between the
two peaks that often decreases with increasing peak frequency,
(5)~have a maximum separation frequency that is close to the spin
frequency inferred from burst brightness oscillations.  A number of
models have been proposed for this phenomenon (see van der Klis 2000 and
references therein).  For the reasons
discussed in Miller (2000), we favor beat-frequency mechanisms, in which
the upper peak frequency is close to the frequency of a circular orbit
at some special radius, and the lower peak is generated by a beat of
that frequency with radiation modulated at the stellar spin frequency
(Miller, Lamb, \& Psaltis 1998).  However, many of the most important
inferences about strong gravity and dense matter from these oscillations
can be derived more generally, from the widely accepted idea that
the upper peak frequency $\nu_2$ is an orbital frequency.

There are two simple considerations that yield a constraint on the
mass $M$ and the radius $R$ of a neutron star, from the observation
of an orbital frequency $\nu_{\rm orb}$.  First, the orbit is obviously
outside the star.  Second, the orbit must also be outside the innermost stable
circular orbit (ISCO), because otherwise there would be rapid inspiral
and any resulting oscillation would last at most a few cycles, leading
to a broad QPO, in conflict with observations.  As shown by Miller et al.
(1998), these limits constrain the mass of the star to be 
$M<2.2(1+0.75j)(1000~{\rm Hz}/\nu_2)\,M_\odot $ and the 
circumferential radius to
be $R<19.5(1+0.2j)(1000~{\rm Hz}/\nu_2)$~km, to first order in the
dimensionless spin parameter $j\equiv cJ/GM^2$, where $J$ is the
stellar angular momentum ($j\approx 0.1$ for a neutron star with a
spin frequency of 300~Hz,
and the spacetime is unique to first order in $j$).
Together, the observation of a given orbital frequency restricts the mass
and radius to be inside a wedge in a $M-R$ diagram; the larger the
frequency, the more restrictive the constraint.
The highest frequency yet observed, 1330~Hz (van Straaten et al. 2000), 
rules out the hard
equation of state L (labeled as in Miller et al. 1998).  This is the
first time that an astrophysical observation has been able to rule out a
hard equation of state.  This is therefore a major advance in our 
understanding of the dense matter in the cores of neutron stars.

As first discussed in 1996, an even more major advance would occur if
strong evidence were found for the existence of the ISCO.  This is because
the existence of unstable circular orbits is a qualitatively new feature
of general relativity compared to Newtonian gravity, and is an essential
component of models of accreting black holes on all scales as well as of
inspiraling compact objects and other phenomena.  
Given that observed frequencies increase with increasing
${\dot M}$, but that $\nu_{\rm ISCO}$ is an upper limit to the frequency
of the upper peak, it was expected that there would be a rollover in the
slope of the $\nu-{\dot M}$ curve for both the upper and lower peaks
(Miller et al. 1998).
Moreover, the frequency at which the curve rolled over has to be a constant,
for many observations of a given source.  This may have been seen in
RXTE data from 4U~1820--30 by Zhang et al. (1998).  If so, this is a
tremendous discovery; not only is this strong confirmation of the prediction
of the ISCO, but it implies a neutron star gravitational mass of
$2.15\,M_\odot$, which limits the possible high-density equation of state
severely.  Given these dramatic implications, care must be taken in the
interpretation of the data.  The main question is whether the rollover 
really happens as a function of ${\dot M}$, or whether the countrate 
change is indicative of a spectral state change.  Further observations and
additional sources with this signature would help immensely.  This is one of 
the many potential benefits of a future large-area timing instrument.

\subsection{Black holes}

There are many different types of quasiperiodic oscillations from 
black hole sources discovered with RXTE.  One that
has been seen in the microquasars GRS~1915+105 and GRO~J1655--40
is particularly intriguing.  The frequency appears not to change
with luminosity in either source (fixed at 67~Hz for GRS~1915+105
and at $\approx$300~Hz for GRO~J1655--40).  In GRS~1915+105 the
feature can be fairly sharp, with a maximum $Q\approx 20$.  These
QPOs are much weaker than many QPOs seen in neutron star sources,
with typical rms amplitudes $\sim$1\%.  They are therefore more
difficult to study, and less is known about them.  However, their
fixed frequency suggests that they are related to some fundamental
quantity or radius in the system; possibilities include frame-dragging
frequencies due to Lense-Thirring precession or periastron advance
(e.g., Stella, Vietri, \& Morsink 1999), and disk oscillation modes
(e.g., Nowak et al. 1997).  Regardless of the physical origin of these
QPOs, the matter generating the oscillations is close enough to the
black hole that it must in some sense reflect the strongly curved
spacetime. It is therefore expected that, especially with the
well-characterized waveforms available with a large-area timing
mission, the mass and spin of the black hole could be inferred, and
potentially the spacetime itself might be probed.

\section{Future Prospects}

An X-ray timing satellite with ten times the collecting area
of RXTE would be the first instrument able to characterize accurately the
waveforms and energy-dependent phase lags
of the high-frequency QPOs observed from compact objects in
low-mass X-ray binaries. As described below, these waveforms and
phase lags encode
otherwise inaccessible information about strong gravity and the
properties of the compact objects. With this capability, such an instrument
would test quantitatively the predictions of general relativity in the
highly curved spacetime near neutron stars and black holes.  It would
also make fundamentally new contributions to our understanding of the
properties of matter at supranuclear densities.

\subsection{Waveforms and Phase Lags}

The waveform from a rotating hot spot during a burst oscillation or
persistent emission is affected by gravitational light deflection and
Doppler shifts.  By fitting the waveform one can therefore constrain
simultaneously the mass and radius of the star, and hence the
equation of state of matter beyond nuclear saturation density, 
which is an issue of central importance
in nuclear physics research. Fig.~1a shows that observations from a
large-area timing instrument would bring an unprecedented level of
precision to our understanding of high-density matter.  In this
simulation we assume a neutron star of gravitational mass 1.8~$M_\odot$
with high-density equation of state UU, and assume that the hot spot is
a small spot on the rotational equator, as seen by a distant observer
in the rotational plane.  We simulate five bursts which had 5\% rms
oscillations at 364~Hz for five seconds each. 
Modern equations of state have mass-radius
curves that span approximately the range between the FPS and UU
equations of state (Akmal, Pandharipande, \&
Ravenhall 1998). It is clear that measurements of this type for several
different neutron stars would exclude most of these EOS. These results
would therefore have direct feedback on nuclear physics.
Similarly, accurate measurement of the waveforms of luminous clumps
around a black hole would yield a precise mass for the hole and strong
constraints on the Kerr spin parameter $j=a/M$.  It may even be possible 
to detect the frequency chirp produced as a
clump moves inside the last stable orbit (Sunyaev 1973, Stoeger 1980).

The high velocity of surface rotation during bursts or persistent
coherent emission also Doppler-boosts the energies of photons from the
hot spot, causing high-energy photons to lead low-energy photons by a
phase that depends on the precise value of the surface velocity and
therefore on the stellar radius, for fixed frequency.   This effect
was first pointed out by Ford (1999) in an analysis of RXTE data on
bursts from Aql~X-1, and has also been analyzed for the coherent
oscillations from the millisecond X-ray pulsar SAX~J1808--3658 by
Weinberg, Miller, \& Lamb (2000) and Ford (2000).  In Fig.~1b we
simulate a fit to such phase lag data, with an assumed gravitational
mass of $2.0\,M_\odot$ and equation of state UU, and phase uncertainties taken
from the SAX~J1808--3658 data.  This figure shows that although RXTE
is not sufficiently sensitive to provide useful constraints, an
observation with a future large-area timing instrument would yield
tight limits on the mass and the radius of this neutron star, and
hence on the equation of state of high-density matter.

\begin{figure}
\plottwo{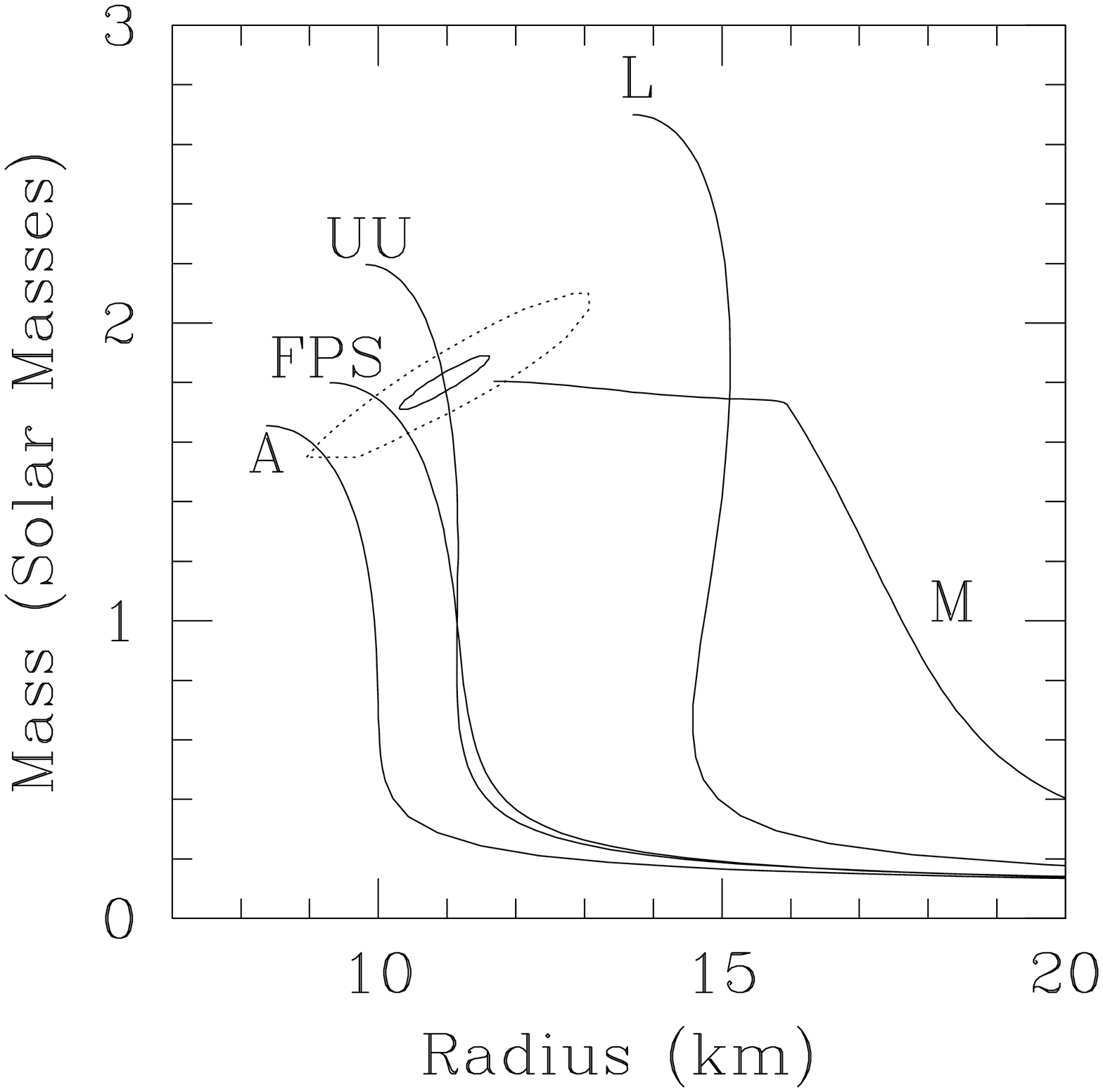}{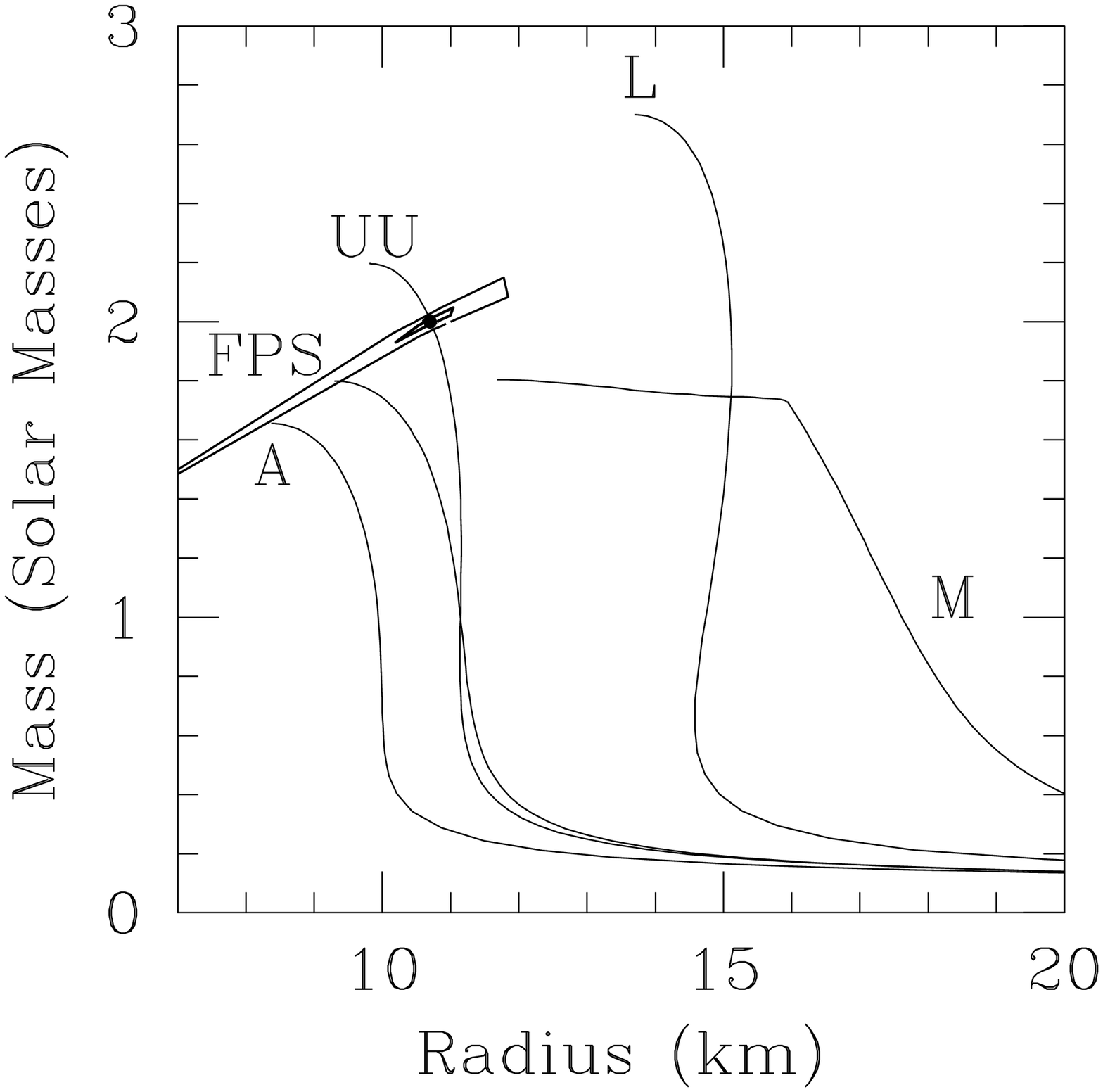}
\caption{(left panel) Constraints on mass and radius possible with 
waveform fitting of burst
brightness oscillations.  We assume the UU high-density equation of
state and a gravitational mass of $1.8\,M_\odot$.  See text for other
details of the simulations.
The dotted outer contour and solid inner contour show, respectively, the
$1\sigma$ confidence region possible with RXTE
and with a future instrument with ten times the area of RXTE. 
The light solid curves show the mass-radius relations
given by different high-density equations of state, labeled as in
Miller, Lamb, \& Psaltis (1998).
(right panel) Constraints on mass and radius from phase lag fitting
of coherent oscillations from a millisecond pulsar.  Phase uncertainties
were adopted from measurements of SAX~J1808--3658, and we assumed 
the UU high-density equation of state and a gravitational mass of
$2.0\,M_\odot$ (indicated by the black dot).  The outer and inner contours
show the 1$\sigma$ uncertainty region possible with, respectively, RXTE and
a future instrument with ten times the area of RXTE.}
\end{figure}

\subsection{Quantitative tests of general relativity}

The most exciting prospect of a high-area timing mission is that
data from it could test quantitatively the predictions of general
relativity in strong gravity.  In the previous sections we have
discussed several independent methods of estimating the mass, radius,
or angular momentum of an individual neutron star or black hole.
These include waveform and phase lag fits and the detection of an ISCO
and the associated orbital frequency.  The combination of all these
independent measurements overdetermines the problem so that the underlying
theories get tested.  For example, the various estimates of the mass
of the central object will agree with each other only if general
relativity is the correct theory of gravity, because general relativity
is assumed in making the estimates.
This would provide a powerful test of its predictions in a regime that is 
otherwise not accessible, and would continue the legacy of X-ray timing
as a clean measuring tool for fundamental physics and astrophysics.

\acknowledgements 
This work was supported in part by NASA grant NAG~5-9756.

\end{document}